\documentclass[12pt]{article}

\begin{document}

\centerline{\Large Sergey P. Novikov\footnote{Math Department and IPST,
University of Maryland, College Park,
MD 20742-2431, USA and Landau Institute for Theoretical Physics, Moscow
117940,
Kosygina 2; phone in USA 301-4054836(o), fax in USA 301-3149363, e-mail
novikov@ipst.umd.edu; this work is supported by the NSF Grant DMS9704613}}

\vspace{0.1cm}

\centerline{\bf 1.Classical and Modern Topology.}

\centerline{\bf 2.Topological Phenomena in Real World Physics}

\vspace{0.1cm}

{\it According to the opinion of the  Ancient Greeks, the famous real and
mythical founders
of Mathematics and Natural Philosophy like Pythagoras, Aristotle
and others, in fact, borrowed them from the Egyptian and Middle
 East civilizations. However, what had been told before in the hidden
  mysteries
 Greek scientists transformed into  written information acceptable for
 everybody. Exactly after that the development of science in the
 modern sense
 started 
 and had already reached a very high level  2000 years ago.
  Therefore you may say that the
 free
 exchange of information and making it clear for people
  have been the most important discoveries of Greeks. I would say it is the
  basis of our science now. As you will see, any violation of this
   fundamental
  rule does serious harm to our science and inevitably leads to its decay.}

\vspace{0.1cm}
\centerline{\bf 1.Classical and Modern Topology.}

{\bf Prehistory. First fifty years of Topology.}
The first important topological ideas were observed
by  famous mathematicians and physicists like Euler, Gauss, Kelvin,
Maxwell and their pupils, during the XVIIIth and XIXth Centuries.
As everybody knows, it was Poincare' who really  started
Topology as a branch of Mathematics in the late XIXth Century.
Many top class mathematicians participated in the development of Topology
in the first half of our century. A huge number of mutually connected
fundamental notions were invented:
 degree of maps and singularities of vector fields,
 homotopy and homology groups,
 differential forms and smooth manifolds,
 the fundamental idea of  transversality,
 the simplicial/cell(CW) and singular 
complexes as  tools for  studying topological invariants, braids,
knot invariants
and 3-manifolds, coverings, fibre bundles and characterisic
classes and many others. Deep connections with Qualitative Analysis,
 Calculus of Variations, Complex Geometry
and Dynamical Systems were established
in this period.  Combinatorial Group Theory and Homological
Calculus started  from  topological sources.
 A great new field of topological
objects unknown to the classical mathematics of the XIXth
 Century appeared finally
in the 1940s. At that time this new area was known to  few number of
 mathematicians only. However there was a very high density of
  really outstanding
 scientists among them.

{\bf 1950s and 60s:  Golden Age of  Classical Topology}.
The fundamental
 set of algebraic ideas unifying all these branches of mathematics
appeared in the 40s; a new era started about 1950.
 Spectral sequences of  fibre bundles, sheafs,
  highly developed homological algebra
of the groups, algebras and modules,
Hopf algebras and coalgebras, were invented and heavily used for
the calculation of  topological invariants needed for the solution
of the fundamental problems of topology. Let me point out that,
 in many cases,
it was
 a completely new type of calculations based on the deep combination of the
very general ''categorial properties'' of these quantities  with
very concrete geometric, algebraic or analytical study
 of a completely new
type.  In the previous period, people
even had no dreams as to  how they could be calculated.
Regular methods were  built
to calculate  homotopy groups, for example.
 It was one of the most difficult
problems of topology. A lot of
 them were computed completely or partially
including the homotopy groups of spheres, Lie groups and homogeneous spaces.
The topologically important cobordism rings were computed and
 used in many
topological investigations.
The famous    signature formula for  differentiable
manifolds was discovered. It has an innumerous number of applications in the
topology of manifolds. Besides that,
 this formula  played a key role in the proof of
the so-called Riemann-Roch theorem in
  Algebraic Geometry and later in the
study index,  the famous homotopy invariant of 
 Fredholm operators.

   The mutual influence of  Topology and Algebraic Geometry
during that period led to the broad extension of the ideas of homology:
the extraordinary (co)homology theories like K-theory and cobordisms appeared.
They brought a new type of technic to topology with many applications.
Representation theory and complex geometry of manifolds deeply unified with
 homological algebra and Hopf algebras. The  technic of formal groups
appeared here. It has been applied in particular for the improvement of the
 calculations of stable homotopy
groups of spheres.
As everybody knows,  during this period topology solved the most
 fundamental
problems in the theory of multidimensional smooth manifolds:

 Nontrivial
differentiable structures on multidimensional spheres were discovered
 on the basis of
the results of algebraic topology combined with a new understanding of
the geometry of manifolds and bundles.
The multidimensional analog of the Poincare Conjecture and
 H-cobordism theorem
were proved.
 Counterexamples to the so-called ''Hauptvermunung der
Topologie'' were found. A classification theory for the multidimensional
smooth (and for PL-manifolds as well) was completely constructed.
 The role of the
fundamental group in this theory led  to the
development of a new branch of algebra:    the algebraic K-theory.
Topological invariance of the most fundamental 
characteristic classes was finally proved.
 The so-called ''Annulus Conjecture''
 was proved.
 No matter how elementary these results can be formulated,
nobody has succeeded to avoid the
use of a whole bunch of results and tools of
 algebraic and differential topology in the proof. The classification theory
for the immersions of  manifolds was constructed. The theory of
 multidimensional
knots was constructed. Several classical problems of the
  theory of 3-manifolds
also were solved during that period: the so-called Dehn's program was
finished after a 50 year  break; the algorithm for
recognizing the trivial knot
in three-space
has been theoretically constructed as a part of the deep understanding of the
structure of 3-manifolds and the surfaces in them.
As a by-product of topology, the fundamental breakthrough in  the topological
understanding of  generic  dynamical systems was reached. A new
great period started in this area.  Qualitative theory of foliations has
 been constructed with especially deep results for 3-manifolds.

As a summary,  I would like to add one more very important characteristic
 of the  topological
community in the golden age of classical topology:

{\bf All important works have been carefully checked. If some theorem
had not been  proved, it immediately became known to everybody}.

So you can find a full set of proofs in the literature. Unfortunately,
a full set of textbooks covering all these developments (1950-1970)
has not been written yet. Many modern textbooks are  written in a very absract way.
Even if they cover some pieces formally, it is more difficult to read them
than the original papers. Let me recommend to you
the Encyclopedia article \cite{N1} written exactly for the exposition
of these ideas. 

{\bf 1970s: Period of decay.}
In  my opinion, the period of the 1970s can be characterized as a period
of decay for  classical topology. There are many indications for that.
Several leading scientists left topology for the new areas like algebra
and number theory, riemannian and symplectic geometry,
dynamical systems and complexity theory,
functional analysis and
 representations, PDEs, and different branches of
 mathematical/theoretical physics....
It is certainly a good characterization of the community if it could generate
such a flux of scientists in many different areas and
bringing to them completely
new ideas. Anyway, this community dispersed.

 What can we say about
the topological community after that?

First of all, some important new ideas appeared           in the 70s
 (like localization technic in
 homotopy topology, the
 nicely organized theory of the rational homotopy type,
hyperbolic topology of 3-manifolds).
However, {\bf a huge informational mess was created in the 1970s}.
Let me point out that a series of fundamental results of that period
was not written, with full proof, until now. Let me give you a list:

Sullivan's Haupvermutung theorem was announced first in early 1967.
 After
the careful analysis made by Bill Browder and myself in Princeton 
of the first version  in May 1967 (before publication),
his theorem was
 corrected: a necessary restriction on the 2-torsion of the group
 $H_3(M,Z)$ was missing. This gap was found and restriction was
 added. Full proof  of this theory  has never been written and published.
  Indeed, nobody
 knows whether it has been finished or not.
 Who knows whether it is complete or not? This question is not clarified
  properly in the literature.
 Many pieces of this theory were developed by other topologists later
 (they used sometimes different ideas). Nobody has unified them until now.
 Indeed, these results were used by many others later.
  In particular, the final Kirby-Siebenmann classification
 of topological multidimensional manifolds therefore is
 not proved yet in the literature.

 The second story is the
  theory of Lipshitz structures on the manifolds.
 In the mid-seventies Sullivan distributed a preprint containing
 the idea how
 to prove existence and uniqueness of such structures on the manifold
 $N^n, n\neq 4$. This idea obviously included (for the uniqueness)
 the direct use of the
   Annulus Conjecture
 (and therefore of  all ideas and technic needed in the
 proof of topological invariance of the
 rational  Pontryagin Classes
 inside).  Proof of the Lipshitz Theory
  has never been published. Indeed, many years later,
 already in the 1990s,
 some brilliant younger scientists developed a very nice
  theory of Fredholm (elliptic)
 operators on Lipshitz manifolds. As a corollary, they claimed that a new
 proof of  topological invariance
  of rational Pontryagin classes has been obtained
 from Analysis (it was a problem posed by Singer in the 60s). 
  Young scientists 
   made a ''logical circle'' believing in the classical results.
    Nobody told them
  that corresponding theorems  have never been proved. How could it happen?
  This funny story shows
  the modern state of information in the topological community.

  Another informational mess has been created in 3D Hyperbolic topology.
  This beautiful area was
  started by Thurston in the mid-70s.
For    many years people could not find out what was proved here. 
  In this area the situation has been finally resolved: it has
   been aknowledged that
these methods lead to the proof of  the original claim
(the so-called Geometrization Conjecture)
only for the special class of Haken manifolds.
The Geometrization Conjecture means more or less that (in the case of closed
3-manifolds) the
fundamental group can be realized as a discrete subgroup acting
in the
3D Hyperbolic space if trivial necessary conditions are satisfied: all its
abelian subgroups are cyclic and $\pi_2=0$. However, it is difficult
to find out
 who actually proved this theorem? It seems for me that the younger
mathematicians who managed to finish this program
 did not receive proper credit.

I would like to mention  that this kind of informational mess
has happened since 1970 not only in topology. For example,
 the famous results
of KAM in the three-body problem known since the early 60s
 were found recently unproved. It was announced for the first time at the
 Berlin Congress last year.
 In this case,
some works supposedly containing full proof were published
 in the first half of
the 60s. Does this mean that nobody accually read them for
at least 30 years?

Do you think that algebra is better? Let me tell you as a curious remark
that  all  works
of the Steklov Institute (i.e. Shafarevich's)
school in algebraic number theory, algebraic geometry
and theory of finite $p$-groups awarded by  the highest (Lenin and State)
prizes
 in the former Soviet Union
since 1959, did not contain full proof. The gaps in the proofs were
 found many years later. Not all these gaps were  really deep.
 However, some of these authors knew their mistakes many years before
 they became publicly known and could not
 correct them.
They managed to fulfill gaps after many years ,
using  much later technical achievements made by other people.
 Does it mean that in the corresponding time, despite 
many public presentations, nobody in fact read  these great works?
Can we say that  all proofs are known now in all these cases? 

There are much worst cases in modern algebra indeed.
 How many of you know that
the so-called classification of  simple finite groups did not
exist as a mathematical theorem until now? In this case we  can even
say that
 in fact (as a few number of real experts have known since 1980)
  no one work existed
 claiming that this problem was finished in this work.
  All public opinion has been based only on the
   ''New York Times Theorem''  for the past 20 years.

    {\bf 1980s and 90s: Period of recovery. The role of Quantum Field Theory.}

   It became clear already in the late 70s that modern quantum
    field theory started
    to generate new ideas in topology. It gave several new alternative ways
     to construct
    topological invariants:  Path integral for the
     metric-independent
    actions on manifolds was used for the first time.
The      famous self-duality equation  appeared
     first 
    in the works of physicists. It was applied in the 80s for the solution of
    fundamental topological problems in the theory of 4-manifolds. Quantum
    string theory
    brought in the
    early 80s new deep results in the theory of the classical Fuchsian groups
    and moduli spaces. At first physicists (like t'Hooft and Polyakov)
    were not interested very much in
    such  by-products of their activity.
     They always said that they were doing
     physics of the real world, not pure mathematics.
      However the next wave of brilliant physicists
     (like  Witten, Wafa and others)
      started to solve problems of pure mathematics.
     Such purely topological subjects like the
     Morse theory  and cobordism theory associated with
     action of compact groups on manifolds,
      were developed in the 80s from the
     completely new point of view.  Symplectic Topology reached a
     very high
     level in the late 80s. We are facing now
      impressive development of Contact Topology.

     Certainly Quantum  Theory  brought  new beautiful ideas. Besides that,
     the fundamental new invariants of knots were discovered in the 80s by
     the topologists who came from  functional analysis and theory
      of $C^*$ algebras.
      These invariants
     also received quantum treatment in the late 80s. The
     beautiful connection of
     the specific Feinmann diagrams with
      surfaces was borrowed from physics literature.
     It 
     became a very effective tool for the solution of
      several topological problems.
     Unfortunately, only
     a few number of mathematicians learned this technic and started to
     apply it in topology.
     I know only  Singer, Konzevich and a very small number of others.
     Even if you will add here the names of pure mathematicians who
      learned this
     with the intention to do real physics,
      this list will increase inessentially. I do not count here
      people who were trained originally
       in the physics community. A
       large number of them moved into  pure mathematics
       with the intention to prove rigorous theorems about
         the models serving
        (in their opinion)
        as an idealization of
        theoretical physics. They call this area Mathematical
         Physics, but not everybody agrees with
        such a definition of mathematical physics.
        This community does not do topology.

        I would like to make a
 remark here concerning
 a beautiful work of Konzevich calculating
certain Chern numbers on the punctured moduli spaces of Riemann Surfaces
 through the
special solution to the KdV hierarchy. This folmula has been
 known as a Witten Conjecture.
 You have to specify for this
some compactification of the moduli spaces of punctured Riemann surfaces,
 otherwise it makes no sense.
Konzevich accually proved this formula for one specific (''Strobel-Penner'')
compactification in 1991. What about
 the standard Deligne-Mumford compactification?
Konzevich claimed in 1992 in his work in Inventiones that it is true.
However, no proof has been presented until now. So this problem is open.
There was a mistakable statement about this at the  Berlin Congress.

Let me point out that the physics community did not create any informational
 mess
in topology. According to their training tradition,  theoretical work
produces  Conjectures  which should be proved only
by some kind of experiment.
Starting to do beautiful nonrigorous mathematics, they do not claim that they
''proved'' something. They are saying that they ''predicted this fact''.
In the case of pure mathematics, the final proof done by pure mathematicians
these people may treat as an ''experimental confirmation''.
In the past ten years several deep results have been obtained
 in the 4D topology.
We cannot say this about 3D topology: quantum invariants here created
 some sort of
''invariantology'':  a lot of people are constructing topological invariants
but no one new topological result has been obtained for almost 10 years.
  Indeed, these ideas look beautiful in some cases. In my opinion,
  new deep results will appear after  better understanding of the relationship of
   new invariants with  classical topology.

\pagebreak

\centerline{\bf Topological Phenomena in  Real World Physics}

\vspace{0.1cm}

{\bf Topological ideas in physics in the period of the early 80s}.
I spent about 10 years learning different parts of 
 Modern Theoretical Physics
in the 60s and 70s. After joining the physics community (i.e. Landau school)
in the early 70s I found out
that most  physicists did not know at all the new areas of mathematics like
topology, dynamical systems and algebraic geometry, including analysis
on Riemann surfaces. The quantum people knew  some extracts from the
group theory and
 representations because they needed it in  Solid State Physics as well as
  in  Elementary
 Particles Theory since the 1960s. A lot of them knew something about
  Riemannian Geometry
 because of the Einsteinian General Relativity.
 However, these people had already  heard something about
the new mathematics of the XXth century and badly wanted to find its
 realization in
 physics. You have to take into account that between them there was a great
 number of extremely
 talented people at that time with very good training
  in  practical
  mathematics.      In some cases I was able to help physicists
  (like Polyakov, Volovic and some others)
 to learn and to use topology in the 70s. I worked this period
  in General Relativity
 (Homogeneous Cosmological Models)
 and Periodic KdV Theory with my pupils and collaborators.
 We found  completely nonstandard applications of Dynamical Systems
 and Algebraic Geometry in these areas. However, until the late 70s I did not
 produce any new topological ideas.
  My very first topological work in physics was
  made in 1980 (see \cite{N2}). I started to use in the spectral theory
 of the Schrodinger operators in periodic lattice and magnetic field
 the idea
  of transversality
 applied to the families of Hermitian matrices or elliptic operators
 on the torus.
 This idea led to the discovery
 of the series of topological invariants, Chern Numbers
 of Dispersion Relations.
 They are well-defined for the generic operators only.
  The classical  Spectral
 Theory in mathematics
 never considered such quantities because they are not defined for every
 operator with prescribed analytical properties of coefficients.
 The ideology of transversality is important here.   
 This work was not understood  by 
 my  colleagues-physicists at that time
 (the vice-editor of JETP did not want to publish it as ''nonphysical'',
 so I published it in the math literature).
 People thought that the important integer-valued
  observable quantities in  Solid State Physics
  may come from  symmetry groups only. Indeed, the Integral Quantum
 Hall phenomenon was discovered soon. Some famous theoretical physicists
  rediscovered
 my mathematical idea after that.
 It is certainly a sum of the Chern classes of dispersion
 relations below the Fermi level.

My next topological discovery was made in the joint work with student
 I.Schmelzer in 1981, dedicated to the very special problem of classical
 mechanics and hydrogynamics (see \cite{NS}).
  I immediately realized its value for  modern
 theoretical  physics, as well as for mathematics, and
 developed this idea in several directions in the same year (\cite{N3}).
 The series of work in the Theory of Normal Metals
 which I am going to discuss today, is also one of  by-products
 of that discovery. Doing the Hamiltonian factorization procedure
  for the top systems
 on the phase spaces like $T^*(SO_3)$ by the action of $S^1$,
  you are coming to
 the systems mathematically equivalent to the motion of the charge particle
 on the 2-sphere. This sphere  is
  equipped by some nontrivial Riemannian metric. What is important and has been
   missed by the good experts in analytical mechanics like Kozlov and Kharlamov
is that  the effective magnetic field
 like Dirac monopole appears here for the nonzero values
 of the ''area integral'' associated with $S^1$-action.
  It means precisely that the magnetic flux along the sphere is nonzero.
The reason for this is that the symplectic (Poisson) structure after
 factorization
is topologically nontrivial. In terms of  modern symplectic geometry,
the magnetic field  is equivalent to
the correction of the symplectic structure.
This fact is not widely known in the
geometric community even now. The appearance of the
topologically nontrivial symplectis structures after $S^1$-factorization
of symplectic manifolds
was independently discovered and formulated
in geometric, nonphysical terminology in 1982
in the beautiful work \cite{DH} for  different goals
 (calculating of integrals).

It has been realized in \cite{NS,N3} that the action functional for such systems
is in fact a closed 1-form on the spaces of loops. These functionals have been
 immediately generalized for higher dimensions, to the spaces of mappings $F$
 of $q$-manifolds in some target space $M$ where a closed $q+1$-form is given
 instead of magnetic field.
 We are coming finally to the action functional well-defined as a closed
 1-form on the mapping spaces $F$. The topological quantization condition for such
 actions was formulated in 1981 \cite{N3,N4} as a condition that this
 closed 1-form should define an integral cohomology
 class in $H^1(F,Z)$. It is necessary
 and sufficient for the Feinmann amplitude to be well-defined as a circle-map

$$\exp\{iS/h\}:F\rightarrow S^1$$

 For the case $q=1$ the original Dirac requirement was
   based on a different idea: the magnetic field
   should be a Chern class for the line bundle whose space of sections
    should serve as
  a Hilbert space of states for our Quantum Mechanics.
  Therefore it should be integral in $H^2(M,Z$).

In  pure topology and in the Calculus of Variations
these ideas led to the
construction of the Morse-type theory for the closed
1-forms on the finite- and
 infinite-dimensional manifolds. Let me refer to the last publication of the
 present author (with P.Grinevich)
 in this direction \cite{NG} where the survey of results and problems
 is discussed. I would like to point out that for the compact symplectic
 manifolds the action functional for any nontrivial Hamiltonian system
 is multivalued. The cohomological class of  symplectic form cannot be
 trivial here. I do not know of such cases in real physics where the
 symplectic
 manifold is compact. However,
 even in the community of symplectic geometers nobody paid attention to
 such properties of action functional
  until the 90s.

 After that I started  to think about different aspects of the
 Hamiltonian Theory where the class of one-valued functions naturally can be
  extended to the
class of all  closed 1-forms. For every symplectic (Poisson)
 manifold $M$ with $H^1(M)\neq 0$ we may consider  Hamiltonian Systems
 generated by the closed 1-form $dH$ where the function $H$ is multivalued.
 Instead of energy levels $H=const$ we have to consider  nontrivial
  codimension 1
 foliation $dH=0$ with Morse (or Morse-Bott) singularities. We are coming
 to the topological problems of studying such foliations. It has been
  posed in \cite{N4}.
 Several participants 
 of my seminar (A.Zorich, Le Tu Thang, L.Alania)
  have made very important contribution to the study of this
  subject.
Interesting quasiperiodic structure appears here. It is not revealed fully
 in my opinion (see references and discussions in the article \cite{N5}).

 {\bf Multivalued Hamiltonians in real physics}.
  I started to look around in 1982 asking the following question:
 can you find such systems in  real physics where 
 Hamiltonian or some other
 important
 integral of motion is multivalued (i.e. $dH$ is  well-defined as a
 closed 1-form)? Much later people realized that in the theory of the
 so-called Landau-Lifshitz
  equation (which is a well-known physical integrable system with
  zero-curvature representation elliptic in the spectral parameter)
 the momentum is a multivalued functional.

At that time (1982) I found only one such system describing 
motion of the quantum  (''Bloch'') electron in the single crystal
D-dimensional normal
metal (D=1,2,3)
under the influence of the homogeneous magnetic field $B$. We are working
here with one-particle approximation for the  system of Fermi particles
whose temperature is low enough. For the zero temperature our electrons
fill in all one-particle quantum  Bloch states $\psi_p$ below the so-called
''Fermi Level'' $\epsilon\leq\epsilon_F$. Its value depends on the number of 
electrons in the system. It is the intrinsic characteristic of our metal.
The index $p$ here may be considered  finally as a point in the torus
$T^D$ defined by the reciprocal lattice dual to the crystallographic one

$$p\in T^D, T^D=R^3/\Gamma^*$$

There is a Morse function $\epsilon(p):T^D\rightarrow R$
 (dispersion relation) such that the domain $\epsilon\leq \epsilon_F$
in the torus $T^D$ is filled in by  Bloch electrons.
Its boundary $\epsilon=\epsilon_F$ is a closed surface $M_F\subset T^D$
for $D=3$. We call it {\bf Fermi Surface.} It is homologous to zero
in the group $H_2(T^3,Z$. For finite but very small temperature
all essential events are happening nearby the Fermi Surface.

Add now a homogeneous magnetic field to our system
(i.e. put metal in the magnetic field $B$). Nobody succeeded in
 constructing
a suitable well-founded theory for the exact description of electrons
in the magnetic field and lattice. Irrational phenomena appear in the
spectral theory of  Schrodinger operators  and destroy
all geometric picture. However, since the late 50s physicists have used
some sort of adiabatic approximation which they call ''semiclassical''.
Let me warn you that this approximation has nothing to do with the
standard
understanding of semiclassical approximation. We take dispersion relation
$\epsilon(p)$ as a function on the torus $T^3$ extracted from the exact
solution of the one-particle Schrodinger operator in the lattice
without magnetic field. We consider a phase space $T^3\times R^3$
with coordinates $p_i,x^j,i,j=1,2,3$ and Poisson bracket of the form:

$$\{x^j,x^k\}=0,\{x^j,p_k\}=\delta^j_k,\{ p_j,p_k\}=B_{jk}$$

\noindent where $B_{jk}(x)$ are  components of the magnetic field $B$
treated as a 2-form. Our space $R^3$ is Euclidean, so we can treat
magnetic field as a vector $B$ with components $B^j$.

Take now the function $\epsilon(p)$ as a Hamiltonian. It generates through
the Poisson structure above a Hamiltonian system in the phase space
$T^3\times R^3$. For the homogeneous (i.e., constant) magnetic field
we can see that our phase space   projects on the torus $T^3$ with
Poisson bracket $\{p_j,p_k\}=B_{jk}$. This Poisson bracket has a Casimir
(Annihilator) $C_B(p)=\epsilon^{ijk}p_iB_{jk}=B^ip_i$.
 This Casimir is multivalued: it is defined by the
  closed 1-form $\omega_B=\sum_iB^idp_i$
on the torus.  As you will see, this is the main reason
for the appearence of  nontrivial topological phenomena in this problem.

Our Hamiltonian $H=\epsilon(p)$
 depends on
the variable $p$ only. Therefore all important information can be extracted from
the Hamiltonian system on the 3-torus with Poisson bracket defined by the
magnetic field. The electron trajectories for the low temperature
can be described as a curves
in this torus such that

$$\epsilon(p)=\epsilon_F, C_B(p)=const$$

However, the levels of the Casimir on Fermi surface are in fact leaves of
foliation given by the closed 1-form restricted on the Fermi surface
$$\omega_B|_{M_F}=\sum_iB^idp_i|_{M_F}=0$$

Some people in ergodic theory studied in fact the most generic ergodic
properties of
''foliations
with transversal measure'' on the Riemann surfaces. In a sense, our situation
is a partial case of that. However, our picture in 3-torus is nongeneric in
that sense. We cannot apply any results of that theory. We have to work
with foliations obtained in the 3-torus by this special procedure only.
 Our use
of word ''generic'' here is resticted by that reqiurement. As we shall see,
 ergodicity is a nongeneric property within this physically realizable
 subclass of foliations 2-surfaces given by the closed 1-form.
 What is interesting is that ergodic examples exist in our picture
 but they occupy a measure zero subset on the sphere of  directions of
 the magnetic fields (if generic Fermi surface is fixed).

As I realized in 1982 (see \cite{N4}), this picture leads to  nontrivial
3-dimensional
topology, and I posed it as a purely topological problem to my students.
 The first beautiful topological
observation was made by A.Zorich \cite{Z1} for the magnetic fields closed
to the rational one. After  new discussion and reconsidering all
 conjectures (see \cite{N5}), I.Dynnikov made a decisive breakthrough in the
 topological understanding of this problem for the generic directions of
 magnetic fields (see \cite{D1}). S.Tsarev constructed
 in 1992
the  first nontrivial ergodic examples, later improved by Dynnikov {see \cite{D2}).

 However, several years passed before some physical results were obtained
 (see the first remark about the possibility of that in
 my article \cite{N6}). We made a series of joint works with A.Maltsev
 (see \cite{NM1,NM2}) dedicated to  physical applications. Essentially,
  we borrowed
 topological results from the works of Zorich and Dynnikov. However, the needs of
 applications required that we not  apply their theorems directly, but
  extract the key points
  from the proofs and reformulate them. So the modern topological
  formulations of these
  results are by-products of these works with applications
   (see the most modern
  survey in \cite{D2}).  Let me formulate here our main physical results
  and after that  explain the topological background and generalizations.

  This picture has been extensively used in  solid state physics since the
   late 50s. The leading theoretical school in that area has been the
    Kharkov-Moscow
   school of I.Lifshitz and his pupils, like M.Azbel, M.Kaganov,
   V.Peschanski, A.Sludskin
   and others. You may find all proper quotations to  physics literature in
   the survey article \cite{NM2}. The
   following fundamental {\bf Geometric Strong
   Magnetic Field Limit} was formulated  by that school (and fully
    accepted later
   by the physics community):

   All essential phenomena in the conductivity of normal metals in
    strong magnetic field
   should follow from the geometry of the dynamical system described above.

   How to understand this principle? You have to take into account that
   this picture certainly will be destroyed by the
    ''very strong'' magnetic field where 
    quantum phenomena (of the magnetic origin) are
    important. It should happen for such magnetic fields that magnetic flux
     through the
     elementary lattice cell is comparable with quantum unit. However, the lattice
     cell in  solid state physics is so small that you need for that
    magnetic field  the order of
    magnitude $B\sim 10^8 Gauss$ or $B\sim 10^4t$
    where 1t is equal to $10^4 Gauss$.

   Therefore we are coming to the conclusion that even for the ''real strong''
    magnetic fields like $10^2t$ this picture still works well.

For our goal we need to consider such metals that Fermi surface is topologically
 nontrivial. It means precisely that the imbedding homomorphism
 of fundamental groups

 $$\pi_1(M_F)\rightarrow \pi_1(T^3)=Z^3$$

\noindent is onto. As people have known already for many years,
the noble metals like copper, gold,
 platinum and others satisfy to this requirement. Probably the very first time
 this property was found was by Pippard in 1956   for copper.
  Many other materials
 with really complicated Fermi surfaces are known now.

By definition, the electron orbit is {\bf compact} if it is periodic
and homotopic to zero
in $T^3$. Therefore it remains compact on the covering surface in $R^3$,
where $R^3$ is a universal covering space
over the torus $T^3$. All other types of trajectories
will be called {\bf noncompact}.

Normally all pictures in
physics literature are drawn in $R^3$, but everybody knows that quasimomentum
vectors $p_1,p_2\in R^3$, such that $p_1-p_2$ belongs to the reciprocal lattice
$\Gamma^*$, are physically identical. 

The  Lifshitz group started to study this dynamical system about 1960
 and made the first important progress.
 For example, Lifshitz and Peschanski
  found some nontrivial examples of noncompact orbits stable under the
variation of the direction of magnetic field.
 It looks like nobody could  understand them properly
in the physics community at that time. It was several decades
 before this community started
to understand the geometry of dynamical systems. The
Lifshitz group was ahead of its time.
They  made some mistakes leading to  wrong conclusions
and investigations were stopped.
You may find the detailed discussion in our survey article \cite{NM2}.
Their mistakes have been found only now because they contradicted  our
final results
describing the conductivity tensor.

{\bf Our main results:}

 Consider projection of the conductivity tensor
on the direction orthogonal to magnetic field.
This is a $2\times 2$ tensor $\sigma_B$. Applying any weak electric field $E$
orthogonal to $B$, we get current $j$.
  Its projection $\sigma_B(E)$
orthogonal to $B$ is only what is interesting for us now. 
We claim that for the strong magnetic field $|B|\rightarrow\infty$
of the generic direction in $S^2$ only two types of asymptotics are possible:

{\bf Topologically Trivial Type}:
 $$\sigma_B\rightarrow 0,|B|\rightarrow\infty$$

More exactly, we have $\sigma_B=O(|B|^{-1})$ for the topologically trivial
type.
All directions with trivial type  occupy a set $U_0$ of measure equal to $\mu_0$
on the two-sphere $U_0\subset S^2$.

{\bf Topologically Nontrivial Type}:
 $$\sigma_B\rightarrow \sigma^0_B+O(|B|^{-1}$$

\noindent Here  $2\times 2$ tensor $\sigma_B^0$ is a nontrivial limit for the
 conductivity
tensor. We claim that it has only one nonzero eigenvalue on the plane
orthogonal to $B$. Let us describe the
 topological properties of this limiting
conductivity tensor. It has exactly one eigen-direction $\eta=\eta_B$ with
eigenvalue equal to zero. Consider  any small variation $B'$
of the  magnetic field $B$. For the new field $B'$ we have
an analogous picture if perturbation is small enough. We have a new $2\times 2$
tensor $\sigma_{B'}$ with one zero eigen-direction $\eta_{B'}=\eta'$.
Our statement is that the plane $a\eta+b\eta', a,b\in R$, generated by this
pair of directions,
is locally stable under the variations of magnetic field. This plane is
integral (i.e. generated by two reciprocal lattice vectors). It contains
zero eigen-directions $\eta_{B''}$ for all small variations of the
magnetic field $B$.
It can be characterized by  3 relatively prime integer numbers
$m=(m_1,m_2m_3)$. This triple of integer
numbers is a measurable topological invariant
of the conductivity tensor. An open set of directions
$U_m\subset S^2$ with measure $\mu_m$ corresponds to this type.
 The total measure of all these types is full:

$$\mu_0+\sum_{m\in Z^3}\mu_m=4\pi$$

We started to look in the old experimental data obtained in
the Kapitza Institute
in the 60-s by Gaidukov and others (see references in \cite{NM2}). They
measured resistance for the single crystal gold samples in the
magnetic field about 2t-4t following the suggestion of Lifshitz.
Confirming the ideas of the Lifshitz group, several domains with
nonisotropic behavior of conductivity were found and
 many suspicious ''black'' dots
(maybe domains of small size) on the sphere $S^2$.  It is not
 hard to see even now that several larger domains in these data
 with nonisotropic conductivity 
should  correspond to the simplest stable topological types like
 $(\pm 1,0,0), (\pm 1,\pm 1, 0), (\pm 1,\pm 1, \pm 1)$ up to permutation
 in the
 natural basis of this cubic lattice. However, for good checking it would be
 nice to increase magnetic field to 20t-40t for a more decisive
 conclusion.  The black dots either correspond to the smaller domains
 with larger values of the topological integers or to some ergodic regimes
 occupying measure zero set on the sphere. For the final decision
 these experiments should be repeated and increased about 10 times
 magnetic field and smaller temperature like $10^{-2}K$.

Let me explain now the topological background of these results.
Consider the generic Morse function $\epsilon: T^3\rightarrow R$
and its generic nonsingular level $M_F\subset T^3, \epsilon=\epsilon_F$
in the torus and in the covering space $M'\subset R^3$. We call the surface
$M'\subset R^3$ {\bf a periodic surface}. Apply now generic magnetic
field $B$ and make the following construction:

Remove all nonsingular compact trajectories (NCT)
from the periodic surface $M'$
and its image $M_F$ in the torus. The remaining part is exactly some surface with
boundary if it is nonempty:

$$M_F \backslash (NCT)=\bigcup M_i$$

\noindent (i.e. Fermi surface minus all NCT is equal to the union of surfaces
 with boundary). We call these surfaces $M_i$ and their closure below
 {\bf the Carriers of Open Trajectories}.
 All boundary curves are the separatrix type trajectories
 homotopic to zero in $T^3$. They bound 2-discs in the corresponding
 planes orthogonal to  magnetic field $B$. Let us fill  them
  by these discs
 in the planes. We get closed piecewise-smooth surfaces $\bar{M}_i$.
 We denote their homological classes by $z_i\in H_2(T^3,Z)$.

 We use the following extract from the proofs of the main theorems of Zorich
 and Dynnikov (see \cite{Z1,D1};
 their theorems have not been formulated  in that way, but
 you may extract these key points from the proofs):

 {\bf In the generic case all these homology classes are nontrivial and
 equal to
  each other up to sign $0\neq z_i=\pm z\in H_2(T^3,Z)$ where $z$ is some
  indivisible
  class in this group.  All these closed surfaces have a genus equal to 1.}

  As you may see, this statement means in fact some kind of the
  ''Topological Complete
  Integrability'' of our systems on the Fermi surfaces
  for the generic magnetic field. 

  For  obtaining our final result on the conductivity tensor, we
  need to use the Kinetic Equation for the quasiparticles based on
   Bloch waves
  nearby the Fermi level. This equation has been used a lot by  solid state
   physicists for the past 30 years. For the small (but nonzero)
   temperature, strong magnetic field and
   apropriate general
   assumptions on the impurities, the motion of quasiparticles
    concentrates along the electron trajectories above. This fact leads to our
     conclusions. Despite  the fact that this theory  is considered  a
     well established one
      already for many years in the physics community, any attempt to
   prove such things as the rigorous
   mathematical theorems would be a huge mess.
    As we see, our final conclusion is separated from
   all theorems by some gap which cannot be eliminated. Let me point out that it
   is always so. ''Rigorous proofs'' in mathematical physics
   never prove anything in  real world physics.

What about nongeneric trajectories? Tsarev and Dynnikov constructed
very interesting examples where genus of carriers of the open trajectories
is larger than 1 (see\cite{D2}). We call such cases {\bf stochastic}.
Sometimes we call them {\bf ergodic}.
There were some attempts to extract from
their properties highly nontrivial asymptotics of the conductivity
tensor in the strong magnetic field \cite{M1}. However, these attempts 
need a better understanding of the properties of  such trajectories.
We have to answer the following questions:

{\bf 1.How many directions of the magnetic field on the sphere $S^2$
admit ergodic trajectories?}

  According to my conjecture, for the generic
Fermi surface, this set of directions has a Hausdorf dimension not greater than
some number  $a<1$ on the sphere $S^2$. For the special Fermi surfaces
$\epsilon=0$ of the even functions like $cos p_1+cos p_2+cos p_3=0$,
we expect to have ergodic trajectories for the set of directions
with Hausdorf dimension like $1<a<2$. Dynnikov started to investigate this
example in his Thesis and proved several general properties.
 Recently R.Deleo investigated such kinds
 of examples more carefully
and performed more detailed calculations (\cite{RL}).
 His results confirm our conjectures. However, the Hausdorf dimension
 of this set has been  unknown in this example until now.

{\bf 2.Which geometric properties  does ''typical'' ergodic trajectory have?}

According to the conjecture of Maltsev, these trajectories are typically
the ''asymptotically self-similar'' plane curves
in the natural sense. His idea (if it is true)
leads to the interesting unusual properties
 of the asymptotic conductivity tensor. Anyway, this problem is very interesting.

Dynnikov investigated also the dependence of these invariants on the level
$\epsilon_F$ of the dispersion relation (see \cite{D2}).  These results are useful
for the 
right understanding of our conjectures.

{\bf Multidimensional Generalizations}.

Consider the following {\bf problem}:
 What can be said about  topology of
the levels $f(x,y)=const$
 of the quasiperiodic functions with $m$ periods on the plane $x,y$?

 For the case $m=3$ this problem exactly coincides with our subject above:
 By definition,  quasiperiodic function on the plane is a restriction
  on the plane $R^2\subset R^m$  of the $m$-periodic function.
  Our space $R^3$ was a space of quasimomenta (more precisely,
  its universal covering). Our
   plane was orthogonal to the magnetic field.
  Can this theory be generalized to the case $m>3$? According to my conjecture,
  it can be generalized  to the case $m=4$. I think that for
   small perturbations of
the rational
  directions this theory can be generalized to any value of $m$.
  We consider now any 4-periodic function $f:R^4\rightarrow T^4\rightarrow R$
  and pair of the
   rational directions $l^0_1,l^0_2$ corresponding to
  some lattice $Z^4$ in $R^4$.

  Let me formulate the following theorem.

  {\bf Theorem}. There exist two nonempty open sets $U_1,U_2$ on the sphere
  $S_3$ containing the rational directions $l^0_1,l^0_2$ correspondingly
   such that:

  {\it For every plane $R^2_l\subset R^4$ from the family 
  given by 2 equations $l_1=const, l_2=const$, the quasiperiodic functions
  $f_l$ have only the following two types of connectivity components of the
  levels $f_l=const$ on the plane $R_l^2$.
 1.The connectivity component of the level is a compact closed curve
    on the plane.
2. The connectivity component of the level is an open curve lying in
the strip of finite width between 2 parallel straight lines with the common
direction
$\eta$. This situation is stable in the following sense.
After any small variations of the directions $l_1\in U_1, l_2\in U_2$,
of function $f$ on $T^4$ or the level we still have such open component
with direction $\eta'$. For all possible perturbations this set of directions
$\eta,\eta',...$ belong to some integral 3-hyperplane in $R^4$.}

This property can be formulated in terms of the integral homology class in
the group $H_3(T^4,Z)$ and of the torical topology of the carriers of the
 open trajectories. The idea of the proof was recently published
 by the author in \cite{N7}.

We may reformulate this problem in terms of Hamiltonian systems.
Let the constant Poisson Bracket $B_{ij}$ be given on the torus $T^m$
whose rank is equal to 2. Any Hamiltonian $f$ generates such systems
whose trajectories are equal to the levels of $f$ on the planes.
Our theorem means that in these cases this Hamiltinian system is Completely
 Integrable in the specific topological sense described above.

\end{document}